\documentclass{JAC2003}

\usepackage{graphicx, subfigure}
\usepackage{booktabs}

\expandafter\def\expandafter\normalsize\expandafter{%
\normalsize\setlength\abovedisplayskip{2pt}}

\expandafter\def\expandafter\normalsize\expandafter{%
\normalsize\setlength\belowdisplayskip{2pt}}

\begin{document}
\title{SIMULATIONS OF COAXIAL WIRE MEASUREMENTS OF THE IMPEDANCE OF ASYMMETRIC STRUCTURES}

\author{H. Day\thanks{hugo.day@hep.manchester.ac.uk}, CERN, Switzerland, Univeristy of Manchester, UK and Cockcroft Institute, UK \\
F. Caspers, E. Metral, CERN, Geneva, Switzerland\\
R.M. Jones, University of Manchester, UK and Cockcroft Institute, UK}

\maketitle

\begin{abstract}
Coaxial wire measurements have provided a simple and effective way to measure the beam coupling impedance of accelerator structures for a number of years. It has been known how to measure the longitudinal and dipolar transverse impedance using one and two wires for some time. Recently the ability to measure the quadrupolar impedance of structures exhibiting top/bottom and left/right symmetry has been demonstrated. A method for measuring the beam coupling impedance of asymmetric structures using displaced single wires and two wire measurements is proposed. Simulations of the measurement system are presented with further work proposed.
\end{abstract}
%
%Introduction
%- Existing wire method to obtain 5 impedances for a structure exhibiting top/bottom, left/right symmetry
%- Previous theoretical work (possibly measure limits of measurement method)
%
%Derivation
%- Start from Tsutsui’s formula from “On single wire measurements of beam coupling impedance”
%- Define the 5 impedances in terms of the generic coefficients
%- Derive the angles necessary to obtain the real and imaginary impedances of all five using single wire method and two wires for dipole impedance
%
%Simulations
%- Geometry with top/bottom, left/right symmetry
%
%- Define geometry (graphite collimator)
%- Define how simulations are done (displaced wire with certain symmmetry planes)
%- Comparison of results with theory (using Tsutsui theory)
%
%- Geometry without top/bottom, left/right
%
%- c-core magnet? With ferrite? or graphite?
%
%Conclusions

\section{INTRODUCTION}
The coaxial wire technique has been used for many years as a bench-top method of measuring the beam coupling impedance of accelerator structures \cite{sands-rees,vaccaro}. There exist single and two-wire measurements techniques to measure the longitudinal and transverse dipolar impedances respectively. Recent work has proposed and utilised a method of measuring the transverse quadrupolar impedance of a structure exhibiting top/bottom, left/right symmetry by combining two wire measurements with those of a displaced single wire \cite{on-swt,ps-mte}. However, many modern accelerator components display little or no symmetry and thus it would be helpful to have a method of measuring the quadrupolar impedance of asymmetric structures. By considering the generic impedance of structure due to a displaced current carrying wire we can determine that a possible method exists and its usefulness is examined through electromagnetic simulations.

\section{IMPEDANCE DUE TO A COAXIAL WIRE IN A GENERIC STRUCTURE}

It is possible to define a generalised longitudinal impedance of a structure caused by an m-th (m=0,1,2,...) order current density $J_{m}$ propogating in the z-direction on a single wire as \cite{on-swt, metral-imp-meet}

\begin{eqnarray}
Z = Z_{0,0} + ae^{-j\theta}(Z_{1,0}+Z_{0,-1}) + ae^{j\theta}(Z_{0,1}+Z_{-1,0}) \nonumber \\ 
+ a^{2}e^{-2j\theta}(Z_{2,0} + Z_{1,-1}+Z_{0,-2}) +a^{2}(Z_{1,1}+Z_{-1,-1}) \nonumber \\
+ a^{2}e^{2j\theta}(Z_{0,2}+Z_{-1,1}+Z_{-2,0}) + O(a^{3})
\label{eqn:longgen}
\end{eqnarray}

where $Z_{m,n}$ (m,n=0,$\pm$1,$\pm$2,...) are the impedances due to the m-th order current density with n-th order azimuthal components, a is the displacement of the wire from the centre axis and $\theta$ is the azimuthal angle of displacement of the wire. It should be noted that $Z_{0,0}$ is what is commonly referred to as the longitudinal impedance, referred to as $Z_{long}$ from hereon, measured at a displacement $a=0$. Subsequently, by using the Panowsky-Wenzel Theorem and using a cartesian coordinate system $x=acos\theta$, $y=asin\theta$ and ignoring constant, coupling and higher order terms, we can define a general transverse impedance in the horizontal and vertical planes

\begin{equation}
Z_{x} = x\left(Z^{dip}_{x} - Z^{quad}   \right),
\end{equation}

\begin{equation}
Z_{y} = y\left(Z^{dip}_{y} + Z^{quad}   \right),
\end{equation}

where x/y is the displacement of the wire in the horizontal/vertical planes respectively,

\begin{equation}
Z^{dip}_{x/y} =\frac{1}{k}\left[ Z_{1,1}\pm Z_{1,-1}\pm Z_{-1,1}+ Z_{-1,-1}\right]
\end{equation}

is the dipolar impedance, $k=\frac{\omega}{c}$ is the wave number, $\omega$ is the angular frequency, $c$ the speed of light and 

\begin{equation}
Z^{quad} = -\frac{2\left(Z_{0,2} +Z_{0,-2}  \right)}{k}
\end{equation}

is the quadrupolar impedance. Here we assume that the source particle and test particle are at the same displacement x/y, as would be the case for a wire measurement, and $\gamma \rightarrow \infty$ such that $Z^{quad}=Z^{quad}_{y}=-Z^{quad}_{x}$. 

The dipolar impedance can be measured directly by the use of a two wire setup, where we measure a longitudinal impedance then normalise by the wave number and wire seperation to obtain the dipolar impedance

\begin{equation}
Z^{dip}_{x/y} = \frac{c Z}{\Delta^{2}_{x/y} \omega}
\end{equation}

where $\Delta_{x/y}$ is the seperation between the two wires.

\section{Wire Method in a Structure with Top/Bottom, Left/Right Symmetry}

\begin{figure}
\begin{center}
\includegraphics[width=0.8\linewidth]{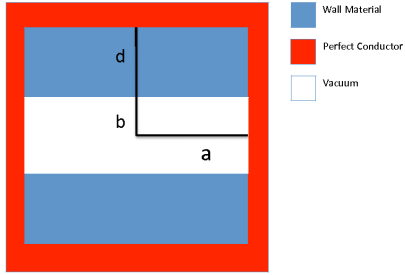}
\end{center}
\caption{An example of a structure with top/bottom, left/right symmetry and the aligned axes of measurement.}
\label{fig:tblrsym}
\end{figure}

If a structure possess top/bottom, left/right symmetry it is possible to define an x- and a y-axis aligned with the lines of symmetry as shown in Fig.~\ref{fig:tblrsym}. It is then possible to greatly simplify the form of Eq.~(\ref{eqn:longgen}) to the following form

\begin{equation}
Z = Z_{long} + k \left[ x^{2}Z_{x} + y^{2}Z_{y}\right].
\end{equation}

It can thus be seen that taking a series of displaced single wire measurements along either the x- or y-axis and fitting a parabola to the resulting measurements gives the total transverse impedance. Coupled with direct measurements of the dipolar impedance using two wires aligned in the correct axis it is thus possible to obtain the longitudinal, dipolar and quadrupolar impedances for a structure.

To test this measurement procedure, simulations of wire measurements of a structure were made using HFSS \cite{hfss-cite}. A structure of the form shown in Fig.~\ref{fig:tblrsym} was chosen to simulate due to analytical models existing to allow a rigorous verification of the method. The simulations are compared to the Tsutsui's formalism \cite{tsutsui-long, tsutsui-dip,tsutsui-quad}. A structure of dimensions a = 25mm, b = 1.5mm, d = 10mm was simulated, with graphite (conductivity $\rho = 7 \times 10^{4}Sm^{-1}$) as the wall material. Displacements were taken at $x=\pm9mm,\pm6mm,\pm3mm,0mm$ and $y=\pm0.7mm,\pm0.5mm,0mm$.

\begin{figure}
\begin{center}
\includegraphics[width=0.8\linewidth]{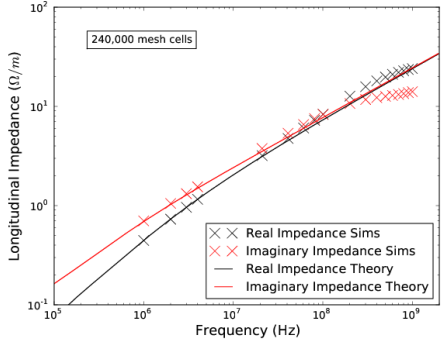}
\end{center}
\label{fig:para-long}
\caption{The longitudinal impedance as simulated using the wire method compared to the Tsutsui theory.}
\end{figure}

\begin{figure}
\begin{center}
\subfigure[]{
\includegraphics[width=0.8\linewidth]{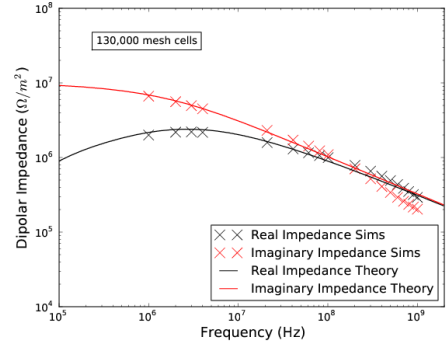}
\label{fig:para-dip-horz}
}

\subfigure[]{
\includegraphics[width=0.8\linewidth]{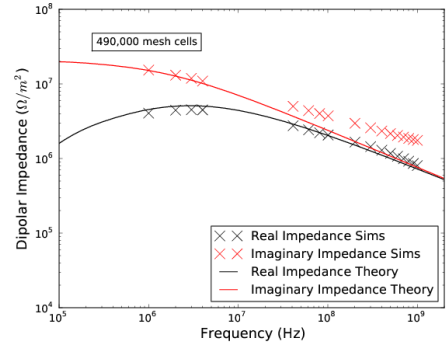}
\label{fig:para-dip-vert}
}
\end{center}
\caption{The a) horizontal and b) vertical dipolar impedance as simulated using the wire method compared to the Tsutsui theory.}
\end{figure}

As we see in Fig.2, \ref{fig:para-dip-horz} and \ref{fig:para-dip-vert} we can replicate the longitudinal and dipolar impedances exceptionally well across the entire frequency range in which the simulation code is suited (above 1MHz into the GHz range). There is some divergence of the imaginary components of the longitudinal and dipolar impedances at high frequencies ($>$600MHz) due to the perturbation of the wire becoming more significant. This is expected to be reduced by using thinner wires radii in the simulations. We can also see in Fig. 4 that the quadrupolar impedance generated by using a single displaced wire and the two wire measurements again agree very well with the Tsutsui's theory, again diverging at higher frequencies due to the perturbation of the wire in the dipolar simulations. Example parabola's are shown in Fig.~\ref{fig:para-real} and \ref{fig:para-imag}.

\begin{figure}
\begin{center}
\includegraphics[width=0.8\linewidth]{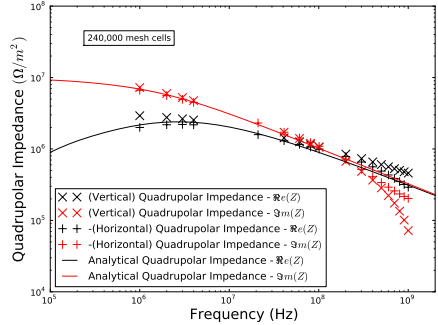}
\end{center}
\label{fig:para-quad}
\caption{The quadrupolar impedances as simulated using the wire method and analysed assuming a top/bottom, left/right symmetric structure.}
\end{figure}

\begin{figure}
\begin{center}
\subfigure[]{
\includegraphics[width=0.45\linewidth]{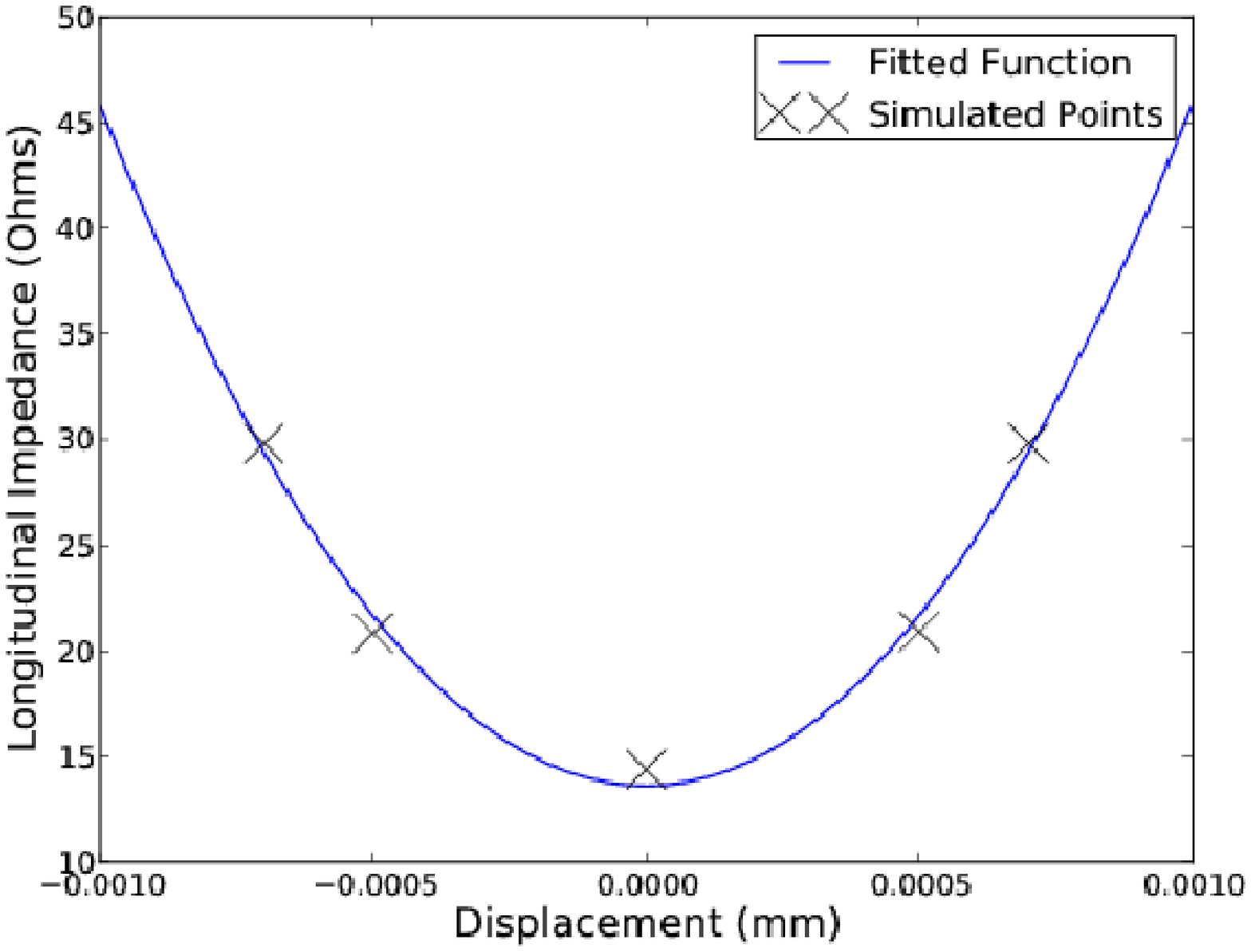}
\label{fig:para-real}
}
\subfigure[]{
\includegraphics[width=0.45\linewidth]{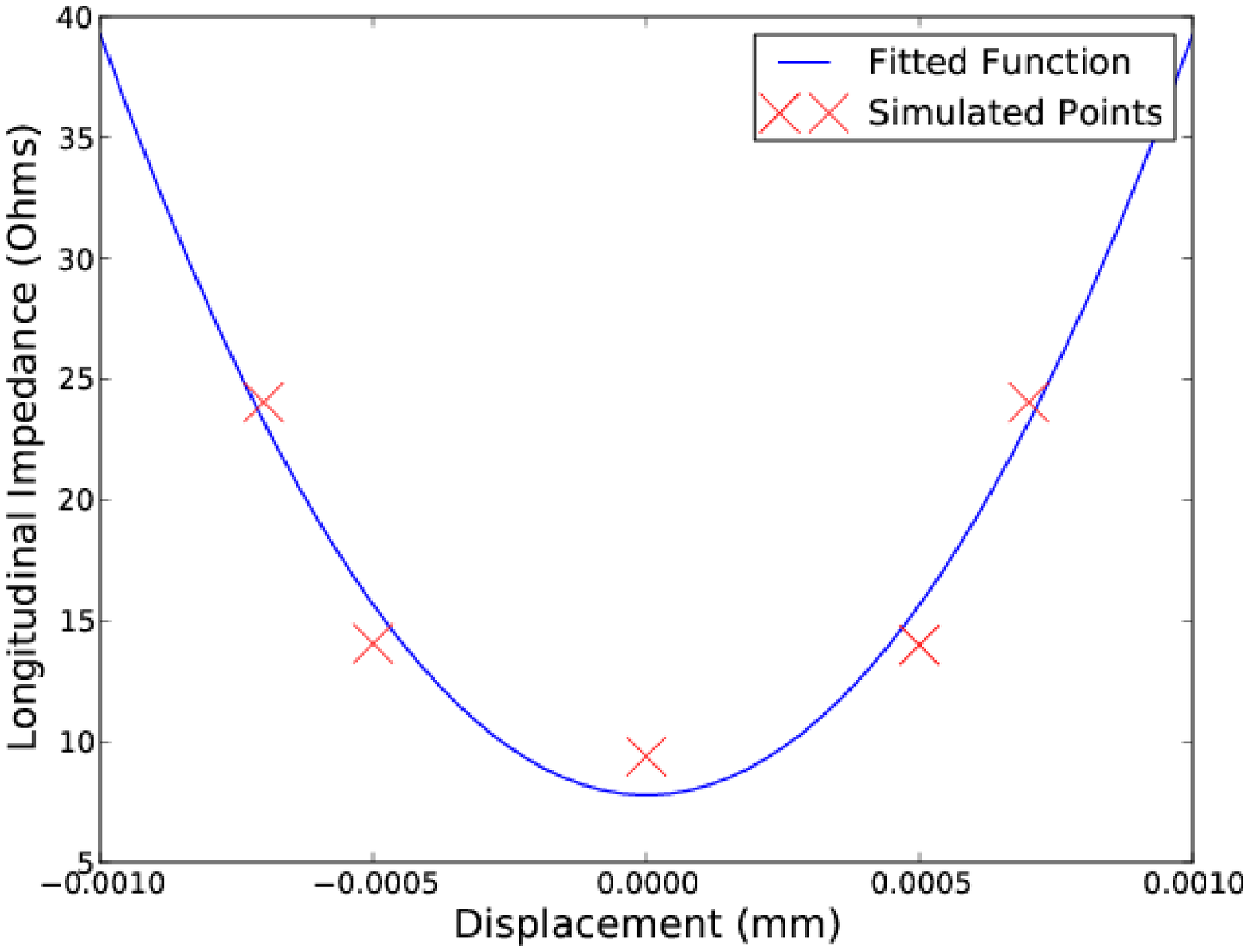}
\label{fig:para-imag}
}
\end{center}
\caption{Example parabola's of the a) real and b) imaginary impedance at 301MHz}
\end{figure}

\section{Wire Method in an Asymmetric Structure}

It can be shown that in the asymmetric case the quadrupolar impedance can be written as

\begin{eqnarray}
Z^{quad} = \frac{Z^{dip}_{x} - Z^{dip}_{y}}{2} \nonumber \\
- \frac{1}{4ka^{2}}[ Z\left(a,\theta = 0\right) + Z\left(a,\theta = \pi\right) \nonumber \\
- Z\left(a,\theta = \frac{\pi}{2}\right) - Z\left(a,\theta = \frac{3\pi}{2}\right)  ].
\end{eqnarray}

This method was analysed by taking the same geometry as before, taking a wire displaced at coordinates (0.5mm,$0$), (0.5mm,$\frac{\pi}{2}$), (0.5mm,$\pi$), (0.5mm,$\frac{3\pi}{2}$). The results are shown in Fig.~6. As can be seen, the accuracy of the results varies heavily depending on the frequency of the simulations. For the real impedance the results match well to the Tsutsui's model 40-200MHz. This is thought to be due to numerical noise being a more substantial problem at these frequencies. Conversely, the imaginary impedance matches well at low frequencies, and begins to diverge more dramatically at higher frequencies. This is due to the innaccurate results of the imaginary impedance of the dipolar measurements. 

\begin{figure}
\begin{center}
\includegraphics[width=0.8\linewidth]{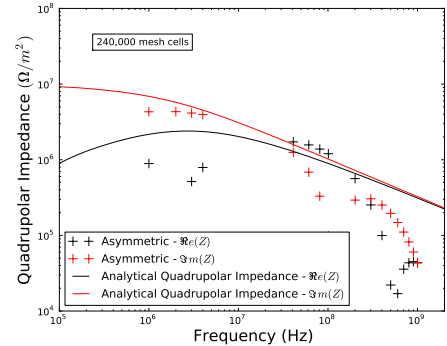}
\end{center}
\label{fig:asym-quad}
\caption{The quadrupolar impedances as simulated using the wire method and analysed assuming an asymmetric structure.}
\end{figure}

\section{CONCLUSION}

We have described a generalised impedance for a particle beam and its equivalent for a coaxial wire simulating beam. We have confirmed existing measurement techniques to allow the measurement of the five impedances most often called for in beam-equipment interactions (longitudinal, horizontal/vertical dipolar, horizontal/vertical quadrupolar) in structures exhibiting top/bottom, left/right symmetry. We have also proposed and simulated a measurement method that allows the similar determination of these impedances in an asymmetric structure. These simulations show some promise, however a number of improvements to these simulations are proposed. Firstly, for low frequencies it would be appropriate to use a code optimised for low frequency simulations. Also alternative methods of calculating the impedance such as using current carrying wires and determining the power loss in the surrounding structure are being investigated which appear promising, but have the disadvantage of only being able to calculate the real impedance. For higher frequencies a denser mesh size is recommended, and thus a smaller geometry has been suggested to allow this with the memory available. When confirmed by successful simulations we hope to verify these using an experimental setup.

\end{document}